\documentstyle[11pt,epsf]{article}

\renewcommand{\thesection}{\Roman{section}.}

\renewcommand{\theequation}{\arabic{section}.\arabic{equation}}

\newcommand{\Eq}[1]{Eq.(\ref{#1})}
\newcommand{\Eqs}[2]{Eqs.(\ref{#1}) and (\ref{#2})}
\newcommand{\Eqss}[3]{Eqs.(\ref{#1}), (\ref{#2}), and (\ref{#3})}

\begin{document}

\title{\bf Short-time scaling behavior of growing interfaces}
\author{Michael Krech \\ Fachbereich Physik, Bergische Universit\"at
 Wuppertal, 42097 Wuppertal \\ Federal Republic of Germany}
\date{}
\maketitle

\begin{abstract}
 The short-time evolution of a growing interface is studied within the
framework of the dynamic renormalization group approach for the
Kadar-Parisi-Zhang (KPZ) equation and for an idealized continuum model
of molecular beam epitaxy (MBE). The scaling behavior of response and
correlation functions is reminiscent of the ``initial slip'' behavior
found in purely dissipative critical relaxation (model A) and critical
relaxation with conserved order parameter (model B), respectively.
Unlike model A the initial slip exponent for the KPZ equation can be
expressed by the dynamical exponent $z$. In 1+1 dimensions, for which
$z$ is known exactly, the analytical theory for the KPZ equation is
confirmed by a Monte-Carlo simulation of a simple ballistic deposition
model. In 2+1 dimensions $z$ is estimated from the short-time
evolution of the correlation function.

\end{abstract}

\newpage

\section{Introduction}
 \setcounter{equation}{0}
Interface formation and growth are typical processes in nonequilibrium
systems. From a technological point of view two important
examples are fluid flow in porous media (oil in rock) \cite{BarStan}
and deposition of atoms during molecular beam epitaxy (MBE)
\cite{BarStan,SLKG}. It is expected that at times much later than
typical aggregation times and on macroscopic length scales these
interfaces develop a characteristic scaling behavior, where the
scaling exponents fall into certain dynamic {\em universality classes}
\cite{BarStan,SLKG,KS} (see below). In certain cases, however,
interfaces can also show turbulent, i.e., spatial multiscaling
behavior \cite{Krug94}. Usually a $d$-dimensional interface is
embedded in $d+1$-dimensional space such that the interface position
at time $t$ can be described by a height function $h({\bf x},t)$,
where ${\bf x}$ denotes the lateral position in a $d$-dimensional
reference plane given by, e.g., the surface of a substrate in MBE.
Complete information about the scaling behavior is contained in the
dynamic structure factor, which is related to the time displaced
height-height correlation function $C({\bf x}-{\bf x'},t,t') \equiv
\langle h({\bf x},t) h({\bf x'},t') \rangle - \langle h({\bf x},t)
\rangle \langle h({\bf x'},t') \rangle$, where a laterally
translational invariant system is assumed. For $t,t' \to \infty$ and
finite $|t-t'|$ the correlation function displays the asymptotic
scaling behavior
\begin{equation} \label{Cscal}
C({\bf x}-{\bf x'},t,t')=|{\bf x}-{\bf x'}|^{2\alpha}
F_C(|t-t'|/|{\bf x}-{\bf x'}|^z),
\end{equation}
where $\alpha$ denotes the {\em roughness} exponent and $z$ is the
{\em dynamic} exponent \cite{BarStan,SLKG}. For a laterally
translational invariant system the interfacial width $w^2(t) \equiv
\langle h^2({\bf x},t) \rangle - \langle h({\bf x},t) \rangle^2$ is
only a function of $t$ and displays the scaling behavior $w(t) \sim
t^\beta$ for late times, where $\beta = \alpha / z$ is the {\em
growth} exponent. For MBE as an example the scaling behavior displayed
in \Eq{Cscal} gives access to the exponents $\alpha$ and $z$ both
experimentally by reflection high energy electron diffraction (RHEED)
(see, e.g., chaper 16 of Ref.\cite{BarStan}) and theoretically by
continuum models \cite{BarStan,SLKG} and Monte-Carlo simulations
\cite{SLKG,PalLan}. Since the advent of the scanning tunneling
microscope (STM) direct imaging techniques for interfaces have also
become an important experimental tool \cite{KHHB}.

Continuum descriptions of interfacial growth processes can be obtained
from general symmetry principles and conservation laws obeyed by the
growth process \cite{BarStan}. The resulting coarse grained growth
model is given by an evolution equation for $h({\bf x},t)$ which has
the form of a Langevin equation with Gaussian distributed noise. This
has first been done in Ref.\cite{EW} for the sedimentation of granular
material and leads to the well known Edwards-Wilkinson (EW) equation.
It is given by
\begin{equation} \label{EWeq}
\textstyle{\partial \over \partial t}h({\bf x},t)
= \nu \nabla^2 h({\bf x},t) + \eta({\bf x},t), 
\end{equation}
where the noise $\eta({\bf x},t)$ has a Gaussian distribution with
$\langle \eta({\bf x},t) \rangle = 0$ and
\begin{equation} \label{noise}
\langle \eta({\bf x},t) \eta({\bf x'},t') \rangle =
2D \delta ({\bf x}-{\bf x'}) \delta (t-t').
\end{equation}
The parameters $\nu$ and $D$ are assumed to be constants and
averages $\langle \dots \rangle$ are taken over the noise
distribution. From explicit solutions of \Eq{EWeq} the exponents
$z$ and $\alpha$ are known exactly in any dimension $d$ of the
interface:
\begin{equation} \label{EWexp}
z = 2 \quad \mbox{and} \quad \alpha = (2 - d)/2 .
\end{equation}
Note that $d = 2$ is the critical dimension of \Eq{EWeq}. One has
$\alpha < 0$ for $d > 2$ so that the height-height correlation
function $C({\bf x}-{\bf x'},t,t')$ according to \Eq{Cscal} {\em
decays} with increasing distance $|{\bf x}-{\bf x'}|$. In $d = 2$
($\alpha = 0$) the correlations increase logarithmically.

The simplest possible nonlinear extension to the EW equation was first
considered systematically in Ref.\cite{KPZ}. The resulting Langevin
equation, usually denoted as the Kadar-Parisi-Zhang (KPZ) equation, is
given by
\begin{equation} \label{KPZeq}
\textstyle{\partial \over \partial t}h({\bf x},t)
= \nu \nabla^2 h({\bf x},t) + \textstyle{\lambda \over 2}
(\nabla h({\bf x},t))^2 + \eta({\bf x},t)
\end{equation}
with Gaussian distributed noise according to \Eq{noise}.
The additional parameter $\lambda$ is again assumed to be a constant.
In the long time limit \Eq{KPZeq} has a global symmetry which is
commonly denoted as Galileian invariance. This invariance originates
from the equivalence of \Eq{KPZeq} to the Burgers equation for a
vorticity-free velocity field ${\bf v}({\bf x},t) = -\nabla
h({\bf x},t)$ and can be stated as follows. If $h({\bf x},t)$
solves \Eq{KPZeq} for some noise function $\eta({\bf x},t)$ then
\begin{equation} \label{Galilei}
h'({\bf x},t) = h({\bf x}-{\bf w}t,t) -\textstyle{1 \over \lambda}
{\bf w}\cdot{\bf x} + \textstyle{1 \over 2\lambda}{\bf w}^2 t
\end{equation}
is a solution of \Eq{KPZeq} for the noise function $\eta'({\bf x},t) =
\eta({\bf x}-{\bf w}t,t)$ and any constant vector ${\bf w}$. An
important consequence is that the exponents $z$ and $\alpha$ of the
KPZ equation fulfill the exact scaling relation \cite{FT,FTH}
\begin{equation} \label{KPZscal}
\alpha + z = 2 .
\end{equation}
Note that in $d = 2$ the EW exponents also obey \Eq{KPZscal}.
The exponents of the KPZ equation are exactly known only in $d = 1$,
where 
\begin{equation} \label{KPZexp}
z = 3/2 \quad \mbox{and} \quad \alpha = 1/2
\end{equation}
due to the existence of a dissipation fluctuation theorem
\cite{FT,DH}. In $d = 2$ numerical investigations indicate $z \simeq 1.6$
and $\alpha \simeq 0.4$ \cite{BarStan}. For $d > 2$ the asymptotic
scaling behavior is either governed by the EW exponents (see
\Eq{EWexp}, weak coupling regime) or by another set of exponents
inaccessible by analytical methods (strong coupling regime) depending
on the value of the effective coupling constant $g \equiv D \lambda^2
/ (4 \nu^3)$ \cite{BarStan,FT,FTH}. In $d = 3$ numerical evidence
suggests $z \simeq 1.7$ and $\alpha \simeq 0.3$ in the strong coupling
regime \cite{BarStan} still indicating rough interfaces in
contrast to EW scaling behavior in $d = 3$ (see \Eq{EWexp}).
Furthermore, it is interesting to note that the nonlinearity in
\Eq{KPZeq} is the most relevant one, i.e., if present it renders all
other nonlinearities irrelevant in the renormalization group sense in
the long-time limit. For intermediate times, however, the presence of
other nonlinearities in the Langevin equation gives rise to various
crossover phenomena \cite{BarStan,SKJJB}. The EW equation and the KPZ
equation for $\lambda \neq 0$ thus represent two different universality
classes for interfacial growth. For $\lambda < 0$ \Eq{KPZeq} can be
viewed as a model for interface {\em corrosion} rather than growth
\cite{KPZ}.

With special regard to MBE growth it is worth noting that
the requirement of mass conservation in ideal MBE \cite{LS}
explicitly excludes the KPZ nonlinearity from a corresponding coarse
grained continuum theory. A simple Langevin equation for ideal MBE has
been proposed in Ref.\cite{LS} (see also Refs.\cite{Sarma94,WV}):
\begin{equation} \label{idMBE}
\textstyle{\partial \over \partial t}h({\bf x},t) = -\nu_1
\nabla^4 h({\bf x},t) + \lambda_1 \nabla^2 (\nabla h({\bf x},t))^2
+ \eta({\bf x},t),
\end{equation}
where $\eta({\bf x},t)$ is chosen according to \Eq{noise}. Mass
conservation in combination with \Eq{noise} immediately leads to the
exact scaling relation $2\alpha - z + d = 0$ for \Eq{idMBE}.
Furthermore, a global symmetry analogous to \Eq{Galilei}, which can be
written in the operator form \cite{SGG}
\begin{equation} \label{GalileiMBE}
{\bf x} \to {\bf x} - 2{\bf w} t \nabla^2 \quad , \quad
h \to h - \textstyle{1 \over \lambda_1} {\bf w} \cdot {\bf x}
\end{equation}
for any infinitesimal vector ${\bf w}$, yields the second exact scaling
relation $\alpha + z = 4$ \cite{LS,SGG}. The exponents $z$ and
$\alpha$ for ideal MBE are therefore known exactly in any dimension of
physical interest:
\begin{equation} \label{idMBEexp}
z = (8 + d)/3 \quad \mbox{and} \quad \alpha = (4 - d)/3
\end{equation}
indicating $d = 4$ as the critical dimension of \Eq{idMBE}.

In this paper \Eqs{KPZeq}{idMBE} are used as paradigms for continuum
descriptions of interfacial growth processes. In linear theory (i.e.
$\lambda = \lambda_1 = 0$) their dynamical exponents are given by
$z = 2$ and $z = 4$ (see \Eqs{EWexp}{MBE0exp}), respectively, and
therefore \Eqs{KPZeq}{idMBE} may be viewed as nonequilibrium
analogues of the dynamical models A and B for critical relaxation,
respectively. In order to invesitgate the scaling behavior of, e.g.,
$C({\bf x}-{\bf x'},t,t')$ for $t' \ll t$ the initial condition
$h({\bf x},t=0)=0$ motivated by deposition processes is used
simultaneously with \Eqs{KPZeq}{noise} or \Eqs{idMBE}{noise},
respectively. Perturbative and nonperturbative aspects of short-time
scaling for the two models are discussed in Secs.II and III within the
framework of dynamic renormalization \cite{JSS,MSR,HH}. Numerical
results from ballistic deposition are presented in Sec.IV and a
summary of the main results is given in Sec.V.

\section{KPZ equation}
 \setcounter{equation}{0}
Due to the spatial translational invariance of the deposition
processes studied here calculations are most conveniently performed in
Fourier space. With the definition $h({\bf x},t) = (2\pi)^{-d} \int d^dq
\exp(i {\bf q} \cdot {\bf x}) h({\bf q},t)$ for the Fourier transform
the dynamic functional ${\cal J}[\tilde{h},h]$ for the KPZ equation
\cite{FT,JSS,Lassig} can be written as the sum of the Gaussian part
\begin{equation} \label{Jhh0}
{\cal J}_0[\tilde{h},h]=\int {d^dq \over (2\pi)^d} \int_0^\infty dt
\left\{ D \tilde{h}({\bf q},t) \tilde{h}(-{\bf q},t)
- \tilde{h}({\bf q},t) \left({\partial \over \partial t}h(-{\bf q},t)
+ \nu {\bf q}^2 h(-{\bf q},t) \right) \right\}
\end{equation}
and the interaction part
\begin{equation} \label{Jhh1}
{\cal J}_1[\tilde{h},h]=-{\lambda \over 2} \int {d^dq_1 \over (2\pi)^d}
\int {d^dq_2 \over (2\pi)^d} \int_0^\infty dt\, {\bf q}_1 \cdot {\bf q}_2\,
\tilde{h}(-{\bf q}_1-{\bf q}_2,t) h({\bf q}_1,t) h({\bf q}_2,t) ,
\end{equation}
where $\tilde{h}({\bf q},t)$ is the Fourier transform of the response
field \cite{MSR}. The initial condition $h({\bf q},0)=0$, which is
implicitly assumed in \Eqs{Jhh0}{Jhh1}, breaks the temporal
translational invariance of the KPZ dynamics. In a more general form
this broken symmetry can be expressed in terms of an additional
contribution to ${\cal J}_0$ which is localized at the time
``surface'' $t = 0$:
\begin{equation} \label{Jhs}
{\cal J}_s[h] = {c \over 2} \int {d^dq \over (2\pi)^d}
(h({\bf q},0) - h_0({\bf q}))^2.
\end{equation}
From the analogy of \Eq{Jhs} with surface contributions to the
Ginzburg Landau functional in the theory of static surface critical
phenomena \cite{HWD} and dimensional arguments the only possible fixed
point values of $c$ under the renormalzation group are $c = \pm
\infty$ and $c = 0$. In the latter case additive renormalizations of
$c$ are supposed to be absorbed in $c$ itself which can be
accomplished by the dimensional regularization scheme. On the other
hand \Eq{Jhs} generates a distribution function $\exp(-{\cal J}_s[h])$
of initial configurations $h({\bf q},0)$ of the deposition field which
leaves the fixed point value $c = \infty$ as the only choice due to
the requirement of normalizability of distribution functions.
Deviations of $c$ from this fixed point value therefore only
generate corrections to scaling \cite{JSS,HWD} which will be
disregarded here. From \Eq{Jhs} one then has the initial condition
$h({\bf q},0) = h_0({\bf q})$. As shown in Appendix A $h_0({\bf q})$
can be incorporated into a source contribution to the dynamic
functional (see \Eqs{Jjj}{jh0q}) and therefore we stick to $h_0({\bf
q}) = 0$ in the following. The correlation and the response propagator
are now easily derived from \Eq{Jhh0}. The results are summarized in
Appendix A.

The introduction of an initial condition striktly speaking also breaks
Galileian invariance (see \Eq{Galilei}). If one demands $h({\bf x},0)
= 0$ as the initial condition for $h$ then $h'({\bf x},t)$ solves
\Eq{KPZeq} with the new initial condition $h'({\bf x},0) =
-{\bf w} \cdot {\bf x} / \lambda$. However, as indicated above one
only has to transform the source fields accordingly in order to
restore the old initial condition. Therefore the Galilei
transformation (see \Eqss{Galilei}{Jjj}{jh0q})
\begin{eqnarray} \label{Galileijj}
h'({\bf q},t) = e^{-i{\bf q}\cdot{\bf w}t} h({\bf q},t) - (2\pi)^d
{i \over \lambda} {\bf w} \cdot {\partial \over \partial {\bf q}}
\delta({\bf q})\, , \quad \tilde{h}'({\bf q},t) =
e^{-i{\bf q}\cdot{\bf w}t} \tilde{h}({\bf q},t)\, , \nonumber \\ \\
j'({\bf q},t) = e^{-i{\bf q}\cdot{\bf w}t} j({\bf q},t)\, , \quad
\tilde{j}'({\bf q},t) = e^{-i{\bf q}\cdot{\bf w}t} \tilde{j}({\bf q},t)
+ (2\pi)^d {i \over \lambda} {\bf w} \cdot
{\partial \over \partial {\bf q}} \delta({\bf q}) \delta(t) \nonumber
\end{eqnarray}
restores the Galileian invariance of the generating functional so that
the corresponding Ward identities (see Ref.\cite{FT}) remain valid.
Note that \Eq{Galileijj} should be read as an infinitesimal
transformation, i.e., terms of order ${\bf w}^2$ have been neglected.

The renormalization group treatment of \Eq{KPZeq} can now be set up
following standard procedures \cite{KPZ,FT,Lassig}. For the case at
hand it is most convenient to combine the dimensional regularization
scheme for the KPZ equation \cite{FT} with the treatment of the
short-time singularites documented in Ref.\cite{JSS}. One defines the
effective coupling constant
\begin{equation} \label{g}
g \equiv D \lambda^2 / (4\nu^3)
\end{equation}
and the renormalized parameters $\nu^R$, $D^R$, and $u$
\cite{KPZ,FT,Lassig}
\begin{equation} \label{Rpar}
\nu^R \equiv Z_\nu \nu\, , \quad D^R \equiv Z_D D\, , \quad \mbox{and}
\quad u \equiv Z_g g\mu^\varepsilon / (2^{d-1} \pi^{d/2} (2-d/2)),
\end{equation}
where $\varepsilon = d-2$ and $\mu$ is an arbitrary momentum scale
which absorbs the naive dimension of $g$ (see \Eq{g}). One finds the
renomalization factors \cite{FT,Lassig} (see also Appendix C)
\begin{eqnarray} \label{Z}
&Z_\nu = 1 + {d-2 \over d}{u \over \varepsilon} + {\cal O}(u^2)\, ,
\quad Z_D = 1 - {u \over \varepsilon} + {\cal O}(u^2) ,& \nonumber \\
&Z_h = \tilde{Z}_h = 1\, ,\quad\mbox{and}\quad Z_g = Z_D Z_\nu^{-3} ,&
\end{eqnarray}
where the $1/\varepsilon$ poles indicate the presence of ultraviolet
singularities \cite{FT}. The nonrenormalization of $h$ and $\tilde{h}$
indicated in \Eq{Z} is {\em exact} and a consequence of \Eq{Gq0} (see
Appendix B). The relation $Z_g = Z_D Z_\nu^{-3}$, which is equivalent
to $\lambda^R = \lambda$, is a consequence of Galileian invariance in
the long-time limit (see \Eq{Galilei} and Refs.\cite{KPZ,FT,Lassig})
and therefore also holds to all orders in perturbation theory. The
renormalization group flow at late times is then governed by the Wilson
functions \cite{FT,Lassig}
\begin{eqnarray} \label{Wilson}
& \zeta_\nu(u) = \textstyle{d-2 \over d} u + {\cal O}(u^2), \quad
\zeta_D(u) = -u + {\cal O}(u^2), & \nonumber \\
& \beta(u) = (d-2 + \zeta_D(u) - 3 \zeta_\nu(u)) u , &
\end{eqnarray}
where the relation between $\beta(u)$, $\zeta_D(u)$, and
$\zeta_\nu(u)$ is again {\em exact}. The higher order corrections to
$\zeta_\nu$ and $\zeta_D$ indicated in \Eq{Wilson} vanish in $d=1$ due
to the existence of a fluctuation-dissipation theorem \cite{FT,DH}. The
fluctuation-dissipation theorem also requires $Z_\nu = Z_D$ in $d = 1$
so that $\zeta_\nu(u) = \zeta_D(u)$ and $\nu/D = \nu^R/D^R$ (see
\Eqs{Rpar}{Z}). We also want to emphasize here that $\zeta_\nu(u)$ and
$\zeta_D(u)$ as given by \Eq{Wilson}, like any other finite order
perturbation theory, do not give access to the strong-coupling regime
of \Eq{KPZeq} for $d \geq 2$.

In analogy with critical phenomena in semiinfinite geometries
\cite{HWD} modifications of the scaling behavior of response and
correlation functions must be expected in the ``time
surface'' $t = 0$ \cite{JSS}. In order to determine the
corresponding anomalous short-time scaling dimensions of response and
correlation functions we introduce two new renormalization factors
$Z_0$ and $\tilde{Z}_0$ by the renormalization prescription (see also
Ref.\cite{JSS})
\begin{equation} \label{Z0}
h({\bf q},0) = Z_0^{1/2} h^R({\bf q},0) \quad \mbox{and} \quad
\tilde{h}({\bf q},0) = \tilde{Z}_0^{1/2} \tilde{h}^R({\bf q},0).
\end{equation}
These $Z$-factors are determined by \Eq{Gq0} and the operator identity
\begin{equation} \label{hth}
{\partial \over \partial t}\, h({\bf q},t=0) = 2D\,
\tilde{h}({\bf q},t=0)
\end{equation}
derived in Appendix B. For the weak coupling regime of the KPZ
equation $(d=1)$ the perturbative analysis of Appendix B consitutes a
rigorous proof of \Eq{hth} and the relations which follow from it (see
below). In the strong coupling regime $(d \geq 2)$, however, the
corresponding perturbative analysis does no longer provide a rigorous
proof of \Eq{hth}, because relations which are valid order by order in
perturbation theory may be violated at a strong coupling fixed point
(see Appendix B). This has to be kept in mind for the following
considerations, although the perturbative result can be regarded as
evidence in favor of the general validity of \Eq{hth}.

From \Eq{Gq0} for $t' = 0$ we immediately find
the exact identity $\tilde{Z}_0 = 1$. Insertion of \Eqs{Rpar}{Z0} into
\Eq{hth} leads to the second exact identity $Z_0 = Z_D^{-2} \tilde{Z}_0$
which determines $Z_0$ in terms of the known $Z$-factor $Z_D$ (see
\Eq{Z}). These identities translate into the {\em exact} relations
\begin{equation} \label{Wilson0}
\tilde{\zeta}_0(u) = 0 \quad \mbox{and} \quad \zeta_0(u) = -2\zeta_D(u)
\end{equation}
among the corresponding Wilson functions (see also \Eq{Wilson}). From
\Eq{Wilson0} one concludes that (i) the response function $G({\bf
q},t,t')$ does not exhibit an anomalous scaling dimension in the
short-time limit $t' \to 0$ (i.e., $t' \ll t$) and that (ii) the
anomalous short-time exponent of the correlation function $C_0({\bf
q},t,t')$ can be expressed by long-time exponents (see \Eq{KPZexp} and
the following text). These properties set KPZ short-time dynamics
markedly apart from model A.

In order to determine the short-time scaling exponent of
$C({\bf q},t,t' \ll t)$ we employ the ``short distance expansion''
\cite{JSS} $h({\bf q},t'\to 0) = \sigma(t') {\partial \over \partial
t'}\, h({\bf q},t'=0) + \dots$ inside the correlation function $C$
which means that
\begin{equation} \label{SDEC}
C({\bf q},t,t' \ll t) = \sigma(t') {\partial \over \partial t'}\,
C({\bf q},t,t'=0) + \dots \, .
\end{equation}
Employing the renormalization prescriptions given by \Eqs{Rpar}{Z0}
one finds
\begin{equation} \label{SDECR}
\sigma(t') = Z_0^{-1/2} \sigma^R(\mu,t',u) \quad \mbox{and} \quad
{\partial \over \partial t'}\, C({\bf q},t,t'=0) = Z_0^{1/2}{\partial
\over \partial t'}\, C^R(\mu,{\bf q},t,t'=0,u)
\end{equation}
for the corresponding renormalized short distance expansion (see
\Eq{SDEC}). Using dimensional analysis the renormalized functions
defined by \Eq{SDECR} can be written in the scaling form
\begin{eqnarray} \label{SDECRscal}
\sigma^R(\mu,t',u) = t' f(y',u) &\quad \mbox{with} \quad&
y'=\nu(\mu) \mu^2 t' \quad \mbox{and} \nonumber \\ \\
{\partial \over \partial t'}\, C^R(\mu,{\bf q},t,t'=0,u) = D(\mu)
g(x,y,u) &\quad \mbox{with} \quad& x = {\bf q}/\mu \quad \mbox{and}
\quad y=\nu(\mu) \mu^2 t, \nonumber
\end{eqnarray}
where $\mu$ has been chosen as the renormalization group flow
parameter. It is now straightforward to derive the renormalization
group equations for the {\em dimensionless} scaling functions
$f(y',u)$ and $g(x,y,u)$ defined by \Eq{SDECRscal}. Using
\Eqs{Rpar}{Z0} one obtains
\begin{eqnarray} \label{RGEfg}
\left[(2+\zeta_\nu(u)) y'{\partial \over \partial y'} + \beta(u)
{\partial \over \partial u} - {\zeta_0(u) \over 2} \right] f(y',u) &=& 0
\quad \mbox{and} \nonumber \\ \\
\left[-x{\partial \over \partial x} + (2+\zeta_\nu(u)) y' {\partial \over
\partial y'} + \beta(u) {\partial \over \partial u} + \zeta_D(u) +
{\zeta_0(u) \over 2} \right] g(x,y,u) &=& 0 . \nonumber
\end{eqnarray}
At the infrared stable renormalization group fixed point $u=u^*$
\Eq{RGEfg} has the solutions
\begin{equation} \label{fgsol}
f(y',u^*) = y'^{\eta_0 / 2z} \quad \mbox{and} \quad
g(x,y,u^*) = y^{-(2\eta_D + \eta_0) / 2z} g'(x^z y),
\end{equation}
where $\eta_a = \zeta_a(u^*)$ for $a=\nu,D,0$, $z = 2+\eta_\nu$, and
$g'$ is a scaling function left undetermined by \Eq{RGEfg}. Combining
\Eqss{SDEC}{SDECRscal}{fgsol} one finds after a few manipulations
\begin{equation} \label{Cqtt}
C({\bf q},t,t' \ll t) = (t'/t)^{1+{\eta_0 \over 2z}}
|{\bf q}|^{\eta_D-z} f_C(|{\bf q}|^z t)
\end{equation}
for the short-time scaling behavior of the correlation function.
For $u^* = 0$ one obtains the EW scaling exponents (see \Eq{EWexp})
and $\eta_0 = 0$ in \Eq{Cqtt}. For any {\em nonzero} fixed point
$u^*$ the exact scaling relation $\eta_D = 3z - 4 - d$ holds (see
\Eq{Wilson}), which is equivalent to \Eq{KPZscal}. From \Eq{Wilson0}
one finally obtains for the short-time exponent (see \Eq{Cqtt})
\begin{equation} \label{theta}
1 + \eta_0/(2z) \equiv \theta = 1 - \eta_D/z = (d+4)/z - 2 .
\end{equation}
In $d=1$ the exact value $\theta = 4/3$ can be obtained from
\Eq{KPZexp}. From numerical estimates for $z$ in $d=2$ and $d=3$ (see
Sec.I) one obtains $\theta \simeq 1.7$ and $\theta \simeq 2.1$,
respectively. The exponent relation given by \Eq{theta} simply means
that the short-time and the long-time scaling behavior of the
correlation function are {\em identical}, i.e., the short-time scaling
behavior can be obtained by extrapolating the $t'$-dependence of
$C({\bf q},t,t')$ from $t' \sim t$ to $t' = 0$. In fact, the scaling
relation given by \Eq{theta} can be derived independently by analyzing
the fluctuation spectrum of the interface displacement velocity
averaged over a macroscopic portion of the interfacial area
\cite{Krug91}.

Finally, we remark that some alternative scaling forms for $C$ can be
obtained from the definition of the growth exponent $\beta = \alpha/z$
which leads to $\theta = d/z + 2\beta$. The scaling behavior displayed
in \Eq{Cqtt} can then be written in the simplified form $C({\bf q},t,t'
\ll t) = t'^\theta g_C(|{\bf q}|^z t)$, where $g_C(y)=y^{-\theta}
f_C(y)$. In real space the correlation function has the scaling form
$C({\bf x},t,t' \ll t) = (t'/t)^\theta |{\bf x}|^{2\alpha}
G_C(t/|{\bf x}|^z)$.

The absence of anomalous scaling exponents for $G({\bf q},t,t')$ for
$t' \ll t$ does not neccessarily mean that $G$ is {\em analytic} for
$t' \to 0$. Exponents describing the asymptotic short-time behavior
are in general functions of the dimensionality $d$ and therefore may
take noninteger values for certain $d$. Similar considerations apply
to the crossover behavior of $G$ for $t \to \infty$ with {\em fixed}
$t-t'$. For details we refer to Appendix C, where some results from
perturbation theory are discussed in the case $d=1$.

\section{Ideal MBE}
 \setcounter{equation}{0}
In terms the deposition field $h({\bf q},t)$ and the response field
$\tilde{h}({\bf q},t)$ the dynamic functional ${\cal J}[\tilde{h},h]$
for \Eq{idMBE} \cite{SLKG,LS,Sarma94} is also written as the sum of
the Gaussian part
\begin{equation} \label{Jhh0MBE}
{\cal J}_0[\tilde{h},h]=\int {d^dq \over (2\pi)^d} \int_0^\infty dt
\left\{ D \tilde{h}({\bf q},t) \tilde{h}(-{\bf q},t)
- \tilde{h}({\bf q},t) \left({\partial \over \partial t}h(-{\bf q},t)
+ \nu_1 ({\bf q}^2)^2 h(-{\bf q},t) \right) \right\}
\end{equation}
and the interaction part
\begin{equation} \label{Jhh1MBE}
{\cal J}_1[\tilde{h},h]=\lambda_1 \int {d^dq_1 \over (2\pi)^d}
\int {d^dq_2 \over (2\pi)^d} \int_0^\infty dt\, ({\bf q}_1 + {\bf q}_2)^2
{\bf q}_1 \cdot {\bf q}_2\,
\tilde{h}(-{\bf q}_1-{\bf q}_2,t) h({\bf q}_1,t) h({\bf q}_2,t) ,
\end{equation}
where the initial condition $h({\bf q},0)=0$ is again implicitly assumed
in \Eqs{Jhh0MBE}{Jhh1MBE}. As described in Sec.II and Appendix A this
special initial condition is sufficient to study the short-time scaling
behavior of response and correlation functions for ideal MBE. The results
of Gaussian theory as implied by \Eq{Jhh0MBE} are summarized in Appendix A.

The further analysis of \Eq{idMBE} can be carried out along the lines
of the analysis of the KPZ equation presented in Sec.II. First, we
note that the invariance under the infinitesimal transformation given
by \Eq{GalileiMBE} in presence of the initial condition $h({\bf
q},0)=0$ is restored by the transformation
\begin{eqnarray} \label{GalileijjMBE}
h'({\bf q},t) = e^{2iq^2{\bf q}\cdot{\bf w}t} h({\bf q},t) - (2\pi)^d
{i \over \lambda_1} {\bf w} \cdot {\partial \over \partial {\bf q}}
\delta({\bf q})\, , \quad \tilde{h}'({\bf q},t) =
e^{2iq^2{\bf q}\cdot{\bf w}t} \tilde{h}({\bf q},t)\, , \nonumber \\ \\
j'({\bf q},t) = e^{2iq^2{\bf q}\cdot{\bf w}t} j({\bf q},t)\, , \quad
\tilde{j}'({\bf q},t) = e^{2iq^2{\bf q}\cdot{\bf w}t} \tilde{j}({\bf q},t)
+ (2\pi)^d {i \over \lambda_1} {\bf w} \cdot
{\partial \over \partial {\bf q}} \delta({\bf q}) \delta(t), \nonumber
\end{eqnarray}
where terms of the order ${\bf w}^2$ have been neglected. In analogy
with the Galileian invariance of \Eq{KPZeq} this symmetry leads to the
nonrenormalization of the nonlinearity: $\lambda_1^R = \lambda_1$ (see
also \Eq{Z} and Refs.\cite{LS,SGG}). Second, \Eq{idMBE} has the global
symmetry of mass conservation which in contrast to \Eq{Rpar}
leads to the {\em additional} nonrenormalization of the noise
correlation amplitude (see \Eq{noise}): $D^R = D$ \cite{LS,SGG}.
If one defines an effective coupling constant by \cite{LS}
\begin{equation} \label{g1}
g_1 \equiv D \lambda_1^2 / \nu_1^3
\end{equation}
and the renormalized parameters $\nu_1^R$, $D^R$, and $u$
\begin{equation} \label{Rpar1}
\nu_1^R \equiv Z_{\nu_1}\nu_1\, , \quad D^R \equiv Z_D D\, , \quad
\mbox{and}\quad u \equiv {Z_{g_1} g_1\mu^\varepsilon \over 2^{d-1} \pi^{d/2}
(2-d/4)}{\Gamma(d/4) \over \Gamma(d/2)},
\end{equation}
where $\varepsilon = d-4$ and $\mu$ is an arbitrary momentum scale
which absorbs the naive dimension of $g_1$ (see \Eq{g1}) then the
renomalization group results for \Eq{idMBE} in the long-time limit can
be summarized as follows (see also Appendix C):
\begin{eqnarray} \label{Z1}
&Z_{\nu_1} = 1 + {d-6 \over d}{u \over \varepsilon} + {\cal O}(u^2)\, ,
\quad Z_D = 1 ,& \nonumber \\
&Z_h = \tilde{Z}_h = 1\, ,\quad\mbox{and}\quad Z_g = Z_{\nu_1}^{-3} .&
\end{eqnarray}
The $1/\varepsilon$ poles indicate the presence of ultraviolet
singularities and the nonrenormalization of $h$ and $\tilde{h}$
indicated in \Eq{Z1} is again a consequence of \Eq{Gq0}. The
corresponding renormalization group flow is therefore governed
by only two nontrivial Wilson functions, namely
\begin{equation} \label{Wilson1}
\zeta_{\nu_1}(u) = \textstyle{d-6 \over d} u + {\cal O}(u^2)
\quad\mbox{and}\quad \beta(u) = (d-4 - 3 \zeta_{\nu_1}(u)) u ,
\end{equation}
where the relation between $\beta(u)$ and $\zeta_{\nu_1}(u)$ is {\em
exact} (see \Eq{Z1}). For any infrared stable fixed point $u^* \neq 0$
\Eq{Wilson1} yields $\zeta_{\nu_1}(u^*) \equiv \eta_{\nu_1} = (d-4)/3$
from which the exponents given by \Eq{idMBEexp} follow directly.

In order to investigate the short-time behavior of the response and the
correlation function of \Eq{idMBE} short-time renormalization factors
$Z_0$ and $\tilde{Z}_0$ are defined as in \Eq{Z0}. From
\Eqs{Gq0}{hth}, which also hold for \Eq{idMBE} (see Appendix B), one
immediately obtains $\tilde{Z}_0 = 1$ and $Z_0 = Z_D^{-2} \tilde{Z}_0
= 1$, where \Eq{Z1} has been used. We thus conclude that in contrast
to KPZ dynamics for ideal MBE {\em neither} the response function
$G({\bf q},t,t')$ {\em nor} the correlation function $C({\bf q},t,t')$
exhibit anomalous scaling behavior for $t' \ll t$ which is
reminiscent of the short-time behavior of model B in critical
relaxation \cite{JSS}. Finally we note that in contrast to model B the
noise in \Eq{idMBE} is {\em not conserved} (see \Eq{noise}).
\Eq{idMBE} with purely {\em conserved} noise has been first considered
in Ref.\cite{SGG} (see also Ref.\cite{BarStan}). The qualitative
short-time behavior is the same as described here. However, with
special regard to MBE the case of purely conserved noise does not play
the same central role as \Eq{idMBE} with nonconserved noise
\cite{SLKG} and we therefore refrain from discussing any details here.

Concerning the asymptotic short time behavior of $G({\bf q},t,t')$ and
$C({\bf q},t,t')$ and the crossover to their asymptotic long-time
behavior one finds properties which are similar to the KPZ behavior
mentioned in Sec.II. Some details obtained from perturbation theory
are reported in Appendix C.

\section{Ballistic deposition}
 \setcounter{equation}{0}
The scaling behavior of $C({\bf q},t,t' \ll t)$ according to \Eq{Cqtt}
can be tested numerically by a Monte-Carlo simulation of a simple
ballistic deposition model on a lattice with periodic boundary
conditions \cite{BarStan}. For convenience we restrict ourselves
to $d=1$ here. The continuum description used in Secs.II and
III is replaced by a discretized description according to
\begin{equation} \label{hxthjn}
h(x,t) = h(x = a j,t = n/(FL)) \equiv a\,h_j(n),
\end{equation}
where the lattice constant $a$ is assumed to be the same both {\em in}
the plane of the substrate and perpendicular to it. The lattice has $L$
sites, $F$ is the incoming particle flux, and $n$ is the number of
deposited particles. Furthermore, the incoming particle flux $F$ has
been normalized to unity, so that $t$ in \Eq{hxthjn} is dimensionless
and given by the number of deposited layers. Finally, $h_j(n)$
defined by \Eq{hxthjn} is also dimensionless and denotes the number
of particles deposited at lattice site $j$ after $n$ particles
have been deposited on the lattice. Ballistic deposition on a
one-dimensional substrate is defined by the {\em deterministic} growth
rule
\begin{equation} \label{bdgr}
h_j(n+1) = \mbox{max}(h_{j-1}(n),h_j(n)+1,h_{j+1}(n))
\end{equation}
(see, e.g., Ref.\cite{BarStan}), where the site $j$ in \Eq{bdgr} has
been selected randomly from the $L$ sites of the lattice. For
periodic boundary conditions $h_1(n)$ and $h_L(n)$ are treated as
nearest neighbors in \Eq{bdgr}.

In order to measure the scaling behavior of $C({\bf q},t,t')$ given by
\Eq{Cqtt} for the above discrete model a discrete Fourier transform
is defined by
\begin{equation} \label{hqn}
\hat{h}_q(n) = {1 \over L} \sum_{j=1}^L h_j(n) e^{-i q a j}
\quad \mbox{with} \quad q = {2 \pi \over L} m ,
\end{equation}
where $m$ is an integer and $0 \leq m \leq L-1$. Using \Eq{hqn}
we define the discrete version of the height-height correlation
function in Fourier space by
\begin{equation} \label{CLqtt}
C_L(q,t,t') = \left\langle \left(\hat{h}_q(n) - \langle
\hat{h}_q(n) \rangle\right) \left(\hat{h}_q(n')-\langle
\hat{h}_q(n') \rangle\right) \right\rangle,
\quad n=FLt, \quad\mbox{and}\quad n'=FLt' ,
\end{equation}
where the triangular brackets $\langle \dots \rangle$ denote an average
over different realizations of the deposition process and the time
arguments $t$ and $t'$ are reintroduced for convenience. For the
measurement of the short-time exponent $\theta$ (see \Eq{theta}) it is
sufficient to measure $C_L(q,t,t')$ for $q = 0$. In this case \Eq{CLqtt}
defines the time displaced correlation function of the spatially
averaged deposition height $\hat{h}_{q=0}(n)$, which can be measured
very quickly during the simulation. In practice a measurement is done
after the deposition of one layer, i.e., the time step is $\Delta t=1$.

\begin{figure}[t]
\centerline{\epsfbox{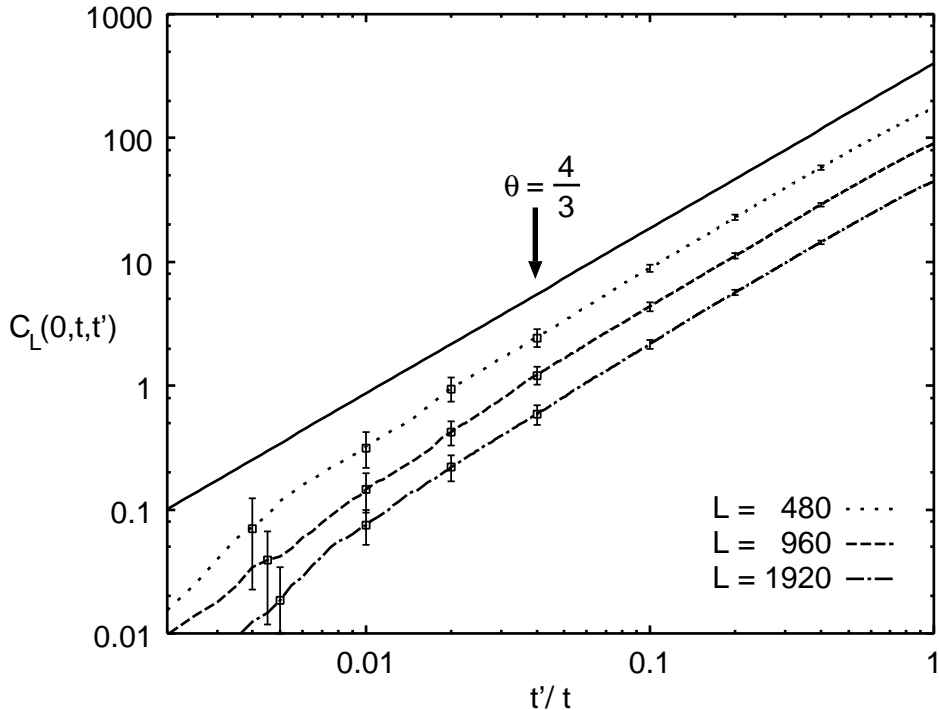}}
\centerline{\parbox{12.5cm}{\caption{\protect\small{
Correlation function $C_L({\bf 0},t,t')$ in $d=1$ as a function
of $t'/t$ for $0.002 \leq t'/t \leq 1$ and $L = 480$ (dotted line), $L
= 960$ (dashed line), and $L = 1920$ (dash-dotted line). The errorbars
are shown only at a few selected points in time and represent one
standard deviation. The solid line displays a power law with the
theoretical short-time exponent $\theta = 4/3$. The data follow this
power law rather accurately in the interval $0.03 \leq t'/t \leq 0.4$
(see main text).}}
\label{CL0tt}}}
\end{figure}
Like a real deposition process the simulation is characterized by an a
priori unknown microscopic aggregation time $t_a$. A scaling behavior
of $C_L$ according to \Eq{Cqtt} can only be observed for $t' \gg t_a$.
On the other hand $t' \ll t$ is required for \Eq{Cqtt} to hold, so
that short-time scaling is restricted to the time window $t_a \ll t'
\ll t$. Furthermore, the lattice size $L$ must be chosen sufficiently
large in order to avoid the onset of finite-size crossover effects if
$t'^{1/z} \sim L$ when $t'$ is still much smaller than $t$. For the
simulation described here $t = 2000$ and $L \geq 480$ fulfill the
above requirements. In order to cope with the very small signal to
noise ratio in each measurement of $C_L(0,t,t')$ for $t' \ll t$
averages are taken over $10^5$ realizations. These are distributed
over 40 individual runs at every point in time for all lattice sizes.
The result is displayed in Fig.\ref{CL0tt}, where $C_L(0,t,t')$ is
shown as a function of $t'/t$ for fixed $t = 2000$ and for $L = 480$,
960, and 1920. For clearity the statistical error is shown only at
a few points in time. As can be seen from Fig.\ref{CL0tt} there is
slightly more than one decade in $t'/t$ available to determine the
short-time exponent $\theta$. Using the least square method in the
interval $0.03 \leq t'/t \leq 0.4$ one finds
\begin{equation} \label{thetabd1}
\theta = 1.349 \pm 0.005
\end{equation}
for $L = 1920$ as the best estimate for $\theta$ from the data shown in
Fig.\ref{CL0tt}. Although the agreement with the theoretical value
$\theta = 4/3$ is very good, there is still a systematic deviation well
outside the statistical error, which is one standard deviation in
\Eq{thetabd1}. One source of systematic errors is the finite lattice
size. For example one finds $\theta = 1.37$ for $L = 480$ and for $L =
240$ (not shown in Fig.\ref{CL0tt}) one even has $\theta = 1.40$ which
indicates that finite lattice corrections to \Eq{Cqtt} are still
visible in \Eq{thetabd1} as a small systematic deviation of $\theta$
from its theoretical value. Furthermore, \Eq{Cqtt} displays only the
{\em leading} singular behavior of the correlation function in the KPZ
universality class. For the ballistic deposition model studied
here corrections to scaling not captured by \Eq{Cqtt} may lead to
sizeable numerical deviations. Therefore, the exponent $\theta$
measured here should be interpreted as an effective exponent. However,
the numerical data for $C_L(0,t,t')$ follow a simple power law
governed by this effective exponent quite accurately. Deviations from
this power law begin to show only for $t'/t > 0.4$, where one is
clearly outside the short-time limit, and for $t'/t < 0.03$, where
microscopic aggregation effects come into play.

In $d = 2$ the ballistic deposition model described here can be used
to estimate the dynamic exponent $z$ of the KPZ equation. The growth
rule for ballistic deposition on a two dimensional square lattice with
$L \times L$ lattice sites is the natural extension of \Eq{bdgr}:
\begin{equation} \label{bdgr2d}
h_{j,k}(n+1) = \mbox{max}(h_{j-1,k}(n), h_{j,k-1}(n), h_{j,k}(n)+1,
h_{j+1,k}(n), h_{j,k+1}(n)),
\end{equation}
where periodic boundary conditions have been assumed. The lattice
momentum has two components and is given by ${\bf q} = (2\pi/L)
(m_1,m_2)$, where $m_1$ and $m_2$ are integers and $0 \leq m_1,m_2
\leq L-1$. The correlation function $C_L({\bf q},t,t')$ is
defined as in \Eq{CLqtt}, where the lattice Fourier transform
$\hat{h}_{\bf q}(n)$ of the deposition field is defined in analogy
with \Eq{hqn}. Note that $n = F L^2 t$ with $F$ normalized to unity
relates $n$ and $t$ in this case so that $t$ is again given by the
number of layers deposited on the substrate. The short-time exponent
$\theta$ can be measured as described above by measuring $C_L({\bf
q}={\bf 0},t,t')$ (see \Eq{CLqtt}) for $t'\ll t$. In order to keep the
amount of CPU time needed for the simulation within reasonable limits
we reduce $t$ to $t = 1000$ and take averages over $2 \times 10^4$
realizations of the deposition process. It turns out that a linear
lattice size of $L = 120$ sites is already sufficient to uniquely
identify at least one decade for the scaling variable $t'/t$ in which
$C_L({\bf 0},t,t')$ obeys the simple power law predicted by \Eq{Cqtt}.
The overall behavior of $C_L({\bf 0},t,t')$ for $L \geq 120$ is
qualitatively the same as displayed in Fig.\ref{CL0tt} so that we
refrain from reproducing it here. For $L = 240$ and $0.01 \leq
t'/t \leq 0.1$ we obtain
\begin{equation} \label{theta2d}
\theta = 1.655 \pm 0.052
\end{equation}
from a least square fit as the best estimate for $\theta$ from the
available data. Using \Eq{theta} we obtain the estimate
\begin{equation} \label{z2d}
z = 1.642 \pm 0.052
\end{equation}
from \Eq{theta2d} as our estimate for the dynamical exponent $z$ in
the KPZ universality class in $d = 2$. A corresponding estimate for
$z$ can be obtained for $L = 120$ which differs by less than half a
standard deviation from the value given by \Eq{z2d}, so that finite
size effects can be neglected within the statistical error. Finally,
we note that according to \Eqs{KPZscal}{z2d} one has $\alpha = 0.358
\pm 0.052$ for the roughness exponent. These values are in agreement
with other numerical data for $z$ and $\alpha$ in $d = 2$ (see chapter
8 of Ref.\cite{BarStan} for a collection of recent estimates) and they
therefore provide some support for the general validity of
\Eqs{hth}{theta}.

\section{Summary and discussion}
 \setcounter{equation}{0}
The following main results have been obtained:
\begin{enumerate}
\item The short-time dynamics of the KPZ equation can be analyzed in
close analogy to the short-time behavior of model A in critical
relaxation. Starting from the operator identity given by \Eq{hth} the
analogy can be summarized schematically:
\begin{equation} \label{modAKPZ}
\begin{array}{ccc}
{\partial \over \partial t}h({\bf q},0) = 2D \tilde{h}({\bf q},0)
& \Longrightarrow & Z_0 Z = Z_D^{-2} \tilde{Z}_0 \tilde{Z} \\ \\
\mbox{model A:} && \mbox{KPZ:} \\
Z_D = (\tilde{Z} / Z)^{1/2} && Z = \tilde{Z} = 1 \\
Z_0 = \tilde{Z}_0 \neq 1 && \tilde{Z}_0 = 1 \, ,\quad Z_0 = Z_D^{-2}.
\end{array}
\end{equation}
In contrast to model A the anomalous short-time scaling dimension
$\theta$ of the deposition field $h$ is given by the dynamical
exponent $z$ (see \Eq{theta}), whereas the response field $\tilde{h}$
does not exhibit an anomalous short-time scaling dimension. The
capability of analytical methods with regard to a full quantitative
description of the crossover behavior from short to long times for the
KPZ equation is limited. A perturbative analysis combined with
dimensional considerations indicate that ${\bf q}^2 (t-t')^2/t^{d/2}$
is the scaling argument which governs the leading finite-time
corrections to the asymptotic long time scaling behavior of the
response function $G({\bf q},t,t')$ in $d=1$. In the correlation
function $C({\bf q}={\bf 0},t,t')$ finite-time corrections
persist indefinitely. A quantitative description of the full scaling
behavior can probably be obtained by combining perturbative methods
with mode coupling theory \cite{FTH}.

\item The short-time dynamics of ideal MBE according to \Eq{idMBE} can
be analyzed in close analogy to the short-time behavior of model B in
critical relaxation. Starting again from \Eq{hth} neither the
deposition field $h$ nor the response field $\tilde{h}$ exhibit
anomalous short-time scaling dimensions, which is the same behavior
as observed for model B \cite{JSS}. In contrast to the KPZ equation
the infrared stable renormalization group fixed point is {\em finite}
in any dimension of physical interest. Therefore purely perturbative
methods can be used to investigate the short-time to long-time
crossover behavior of the response and the correlation function
within, e.g., an $\varepsilon = d-4$ expansion. In combination with
dimensional arguments the perturbative analysis indicates that the
scaling argument $q^4 (t-t')^2/t^{d/4}$ governs the leading
finite-time correction to the asymptotic long-time behavior of the
response function $G({\bf q},t,t')$. In the correlation function
$C({\bf q}={\bf 0},t,t')$ finite-time corrections again persist
indefinitely.

\item With a simple ballistic deposition model \Eq{theta} can be used
to measure the dynamical exponent $z$ for the KPZ universality class
from a simulation of the short-time behavior of the height - height
correlation function. Although such a simulation in principle requires
short computer times, the overall benefit is somewhat limited due to
the small signal to noise ratio in the correlations for $t' \ll t$
(see Fig.\ref{CL0tt}) which in turn must be compensated for by running
the simulation with high statistics. In $d=1$, where \Eq{KPZexp} gives
the exact scaling exponents, the numerical results for $\theta$ agree
very well with the theoretical value $\theta = 4/3$ which is
equivalent to $z = 3/2$. In $d=2$ \Eq{theta} has been successfully
used to obtain a numerical estimate for the dynamical exponent $z$ in
the KPZ universality class (see \Eq{z2d}).
\end{enumerate}

Finally, it should me mentioned that the short-time scaling behavior
of the magnetization in an Ising model with model A (Glauber) dynamics
can be efficiently used to determine the dynamic and static critical
exponents in the Ising universality class \cite{LSZ}. It would be
interesting to see, to what extent Monte-Carlo methods similar to
those described here and in Ref.\cite{LSZ} can be used to study the
asymptotic long-time scaling behavior of interfacial growth models
from their short-time dynamics. The scaling relation between $\theta$
and $z$ may also open an alternative path for direct numerical
investigations of the KPZ equation.

\noindent
{\Large{\bf Acknowledgment}}

The author gratefully acknowledges useful correspondence with S. Pal,
J. Krug, H.W. Diehl, and D.P. Landau.

\setcounter{section}{0}
\renewcommand{\thesection}{Appendix \Alph{section}:}
\renewcommand{\theequation}{\Alph{section}.\arabic{equation}}

\section{Gaussian theory}
 \setcounter{equation}{0}
The Gaussian part ${\cal J}_0$ of the dynamic functional for \Eq{KPZeq}
is the same as for model A of critical relaxation \cite{JSS} and can
be written in the symmetric form
\begin{equation} \label{Jhh0sym}
{\cal J}_0[\tilde{h},h] = {1 \over 2} \int {d^dq \over (2\pi)^d}
\int_0^\infty dt\, \left( \tilde{h}(-{\bf q},t), h(-{\bf q},t) \right)
{\cal A} \left(\begin{array}{c} \tilde{h}({\bf q},t) \\ h({\bf q},t)
\end{array} \right),
\end{equation}
where $h({\bf q},0) = 0$ and the response field fulfills the additional
condition $\tilde{h}({\bf q},\infty) = 0$. The self-adjoint matrix
operator $\cal A$ is then given by
\begin{equation} \label{A}
{\cal A} = \left(\begin{array}{cc} 2D & -{\partial \over \partial t} -
\nu {\bf q}^2 \\ {\partial \over \partial t} - \nu {\bf q}^2 & 0
\end{array} \right).
\end{equation}
In terms of the source fields $\tilde{j}$ and $j$ introduced by adding the
source term
\begin{equation} \label{Jjj}
{\cal J}_j[\tilde{h},h] = \int {d^dq \over (2\pi)^d} \int_0^\infty dt\,
\left(
\tilde{h}({\bf q},t) \tilde{j}(-{\bf q},t) + h({\bf q},t) j(-{\bf q},t)
\right)
\end{equation}
to \Eq{Jhh0sym} the generating functional
\begin{equation} \label{Wjj0}
{\cal W}_0[\tilde{j},j] = \ln \int {\cal D} \tilde{h} \int {\cal D} h
\exp \left\{{\cal J}_0[\tilde{h},h] + {\cal J}_j[\tilde{h},h] \right\}
\end{equation}
is conveniently evaluated by solving the set of initial value problems
given by
\begin{eqnarray} \label{DEQhh}
2D\tilde{h}({\bf q},t) - \left({\partial \over \partial t} + \nu
{\bf q}^2 \right) h({\bf q},t) + \tilde{j}({\bf q},t) = 0
&\, ;&\, h({\bf q},t) = 0 \nonumber \\ \\
\left({\partial \over \partial t} - \nu {\bf q}^2 \right)
\tilde{h}({\bf q},t) + j({\bf q},t) = 0
&\, ;&\, \tilde{h}({\bf q},\infty) = 0 \nonumber
\end{eqnarray}
for $\tilde{h}$ and $h$. The more general initial condition $h({\bf
q},0) = h_0({\bf q})$ can be incorporated in the source field
$\tilde{j}({\bf q},t)$ by the replacement
\begin{equation} \label{jh0q}
\tilde{j}({\bf q},t) \to \tilde{j}({\bf q},t) + \delta(t) h_0({\bf q}).
\end{equation}
The solution of \Eq{DEQhh}, which is equivalent to calculating the
inverse of the operator $\cal A$ (see \Eq{A}), is given by
\begin{equation} \label{Solhh}
\left( \begin{array}{c} \tilde{h}({\bf q},t) \\ h({\bf q},t)
\end{array} \right) = \int_0^\infty dt' \left( \begin{array}{cc}
0 & G_0({\bf q},t',t) \\ G_0({\bf q},t,t') & C_0({\bf q},t,t') \end{array}
\right) \left( \begin{array}{c} \tilde{j}({\bf q},t') \\ j({\bf q},t')
\end{array} \right),
\end{equation}
where
\begin{eqnarray} \label{GC}
G_0({\bf q},t,t') &=& \Theta (t-t') e^{-\nu {\bf q}^2 (t-t')} \quad
\mbox{and} \nonumber \\ \\
C_0({\bf q},t,t') &=& {D \over \nu {\bf q}^2} \left(
e^{-\nu {\bf q}^2 |t-t'|} - e^{-\nu {\bf q}^2 (t+t')} \right)
\nonumber
\end{eqnarray}
are the response and correlation functions of Gaussian theory for the
KPZ equation, respectively. From \Eqs{Wjj0}{GC} one obtains for the
generating functional
\begin{equation} \label{Wjj0GC}
{\cal W}_0[\tilde{j},j] = \int {d^dq \over (2\pi)^d} \int_0^\infty dt
\int_0^\infty dt' \left(j(-{\bf q},t) G_0({\bf q},t,t') \tilde{j}({\bf q},t') 
+{1 \over 2} j(-{\bf q},t) C_0({\bf q},t,t') j({\bf q},t') \right).
\end{equation}
For the general initial condition $h({\bf q},0) = h_0({\bf q})$ the
corresponding generating functional is obtained by applying the
replacement \Eq{jh0q} directly to \Eq{Wjj0GC}.
The response and correlation propagators can now be obtained
by functional derivatives of \Eq{Wjj0GC} with respect to $\tilde{j}$
and $j$:
\begin{eqnarray} \label{GpCp}
{\cal G}_0({\bf q},t;{\bf q}',t') \equiv \langle h({\bf q},t)
\tilde{h}({\bf q}',t') \rangle_0 &=& (2\pi)^d \delta ({\bf q}+{\bf q}') 
G_0({\bf q},t,t') \nonumber \\ \\
{\cal C}_0({\bf q},t;{\bf q}',t') \equiv \langle h({\bf q},t)
h({\bf q}',t') \rangle_0 &=& (2\pi)^d \delta ({\bf q}+{\bf q}') 
C_0({\bf q},t,t'), \nonumber
\end{eqnarray}
where $\langle \dots \rangle_0$ denote the average with respect to the
Gaussian distribution generated by \Eq{Jhh0sym}. From momentum
conservation it is obvious that the {\em full} two-point correlation
functions ${\cal G}({\bf q},t;{\bf q}',t')$ and ${\cal C}({\bf
q},t;{\bf q}',t')$ can be written in the same form as their gaussian
counterparts (see \Eq{GpCp}) which serves as the definition of the
full response function $G({\bf q},t,t')$ and the full correlation
function $C({\bf q},t,t')$. One should also note that
the simultaneous requirements $h({\bf q},0) = 0$ and $\tilde{h}({\bf
q},\infty) = 0$ forbid a Fourier transformation with respect to time
so that one has to stick to the above mixed representation of the
propagators for further calculations. Especially the normalization
conditions imposed on correlation functions in order to define {\em
renormalized} quantities (see \Eqs{Rpar}{Z}) have to be reformulated
accordingly. Note that the exponents $\alpha$ and $z$ implied by
\Eq{GC} are the Edwards-Wilkinson exponents given by \Eq{EWexp}.

In close analogy to the considerations described above
the Gaussian part of the dynamic functional for \Eq{idMBE} is the same
as for model B of critical relaxation \cite{JSS,HH} and can be written
in the same symmetric form as given by \Eq{Jhh0sym} together with the
conditions $h({\bf q},0) = 0$ and $\tilde{h}({\bf q},\infty) = 0$.
In this case the self-adjoint matrix operator $\cal A$ is given by
\begin{equation} \label{AMBE}
{\cal A} = \left(\begin{array}{cc} 2D & -{\partial \over \partial t} -
\nu_1 q^4 \\ {\partial \over \partial t} - \nu_1 q^4 & 0
\end{array} \right),
\end{equation}
where $q = |{\bf q}|$ is the modulus of the momentum vector ${\bf q}$.
The generating functional given by \Eq{Wjj0} is evaluated by solving
the corresponding initial value problem for $\tilde{h}({\bf q},t)$ and
$h({\bf q},t)$ (see \Eq{DEQhh}). The solution can be written in the
same form as \Eq{Solhh}, where instead of \Eq{GC} one has
\begin{eqnarray} \label{GCMBE}
G_0({\bf q},t,t') &=& \Theta (t-t') e^{-\nu_1 q^4 (t-t')} \quad
\mbox{and} \nonumber \\ \\
C_0({\bf q},t,t') &=& {D \over \nu_1 q^4} \left(
e^{-\nu_1 q^4 |t-t'|} - e^{-\nu_1 q^4 (t+t')} \right)
\nonumber
\end{eqnarray}
for the response and the correlation function, respectively, of Gaussian
theory for \Eq{idMBE}. With $G_0$ and $C_0$ taken from \Eq{GCMBE} the
corresponding response and correlation propagators are again given by
\Eq{GpCp}. We close this section by noting that the exponents $\alpha$
and $z$ implied by \Eq{GCMBE} are given by
\begin{equation} \label{MBE0exp}
z = 4 \quad \mbox{and} \quad \alpha = (4 - d)/2
\end{equation}
in contrast to \Eq{idMBEexp}.

\section{Perturbation theory}
 \setcounter{equation}{0}
Due to the presence of strong coupling fixed points in the KPZ equation
for $d \geq 2$ perturbation theory is only of limited value as compared
to perturbation theory for model A critical dynamics, for example.
However, some rigorous relations can be proved by analyzing the building
blocks of perturbation theory for response and correlation functions
and therefore some details concerning perturbative calculations for
\Eqs{KPZeq}{idMBE} will be described below.

\begin{figure}[t]
\centerline{\epsfbox{gcv.eps}}
\centerline{\parbox{12.0cm}{\caption{\protect\small{
(a) Graphical representation of the response propagator ${\cal G}_0$
and the correlation propagator ${\cal C}_0$ (see \protect\Eq{GpCp}).
(b) Graphical representation of the vertex of \protect\Eq{KPZeq}
(see \protect\Eq{Jhh1}). The wiggly lines represent the response field
$\protect\tilde{h}({\bf q},t)$ and the straight lines represent the
deposition field $h({\bf q},t)$.}}
\label{GCV}}}
\end{figure}
\begin{figure}[t]
\centerline{\epsfbox{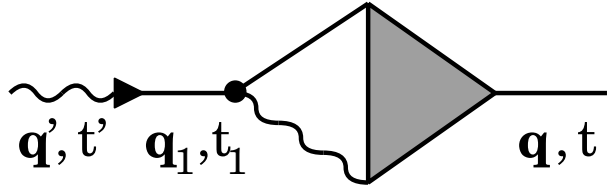}}
\centerline{\parbox{12.0cm}{\caption{\protect\small{
Block diagram for the response function $G({\bf q},t,t')$. The shaded
triangle consists of an arbitrary number of vertices and propagators.
To lowest order it is given by the vertex displayed in
Fig.\protect\ref{GCV}(b) (see main text).}}
\label{Gblck}}}
\end{figure}

For the response and correlation propagators given by \Eq{GpCp} we use
the graphical representation shown in Fig.\ref{GCV}(a). The vertex and
its analytical expression can be read off from \Eq{Jhh1}, they are
shown in Fig.\ref{GCV}(b). The momentum carried by the response field
in Fig.\ref{GCV}(b) is $-{\bf q}_1-{\bf q}_2$. Contributions to
response, correlation, and vertex functions can be constructed from
the elements in Fig.\ref{GCV} according to the standard Feynman rules
of dynamic perturbation theory \cite{FT,JSS,HH}. As a first example we
analyze the response function $G({\bf q},t,t')$. Any contribution to
$G$ from a perturbation expansion can be cast into the form of the
block diagram shown in Fig.\ref{Gblck}. 
According to the Feyman rules the first vertex contribution to an
arbitrary diagram for $G$ has to be arranged as shown in
Fig.\ref{Gblck}. The remainder of the diagram, which is not
neccessarily one-particle irreducible, is indicated by the shaded
triangle and may be interpreted as an arbitrary contribution to the
three-point vertex function. To lowest order this three-point vertex
function is shown in Fig.\ref{GCV}(b). From the explicit momentum
dependence of the vertex it is obvious, that for zero momentum ${\bf
q}'={\bf q}$ the block diagram displayed in Fig.\ref{Gblck} vanishes
identically. One therefore finds the exact relation
\begin{equation} \label{Gq0}
G({\bf q}={\bf 0},t,t') = G_0({\bf q}={\bf 0},t,t') = \Theta(t-t')
\end{equation}
for the response function of \Eq{KPZeq}.

\begin{figure}[t]
\centerline{\epsfbox{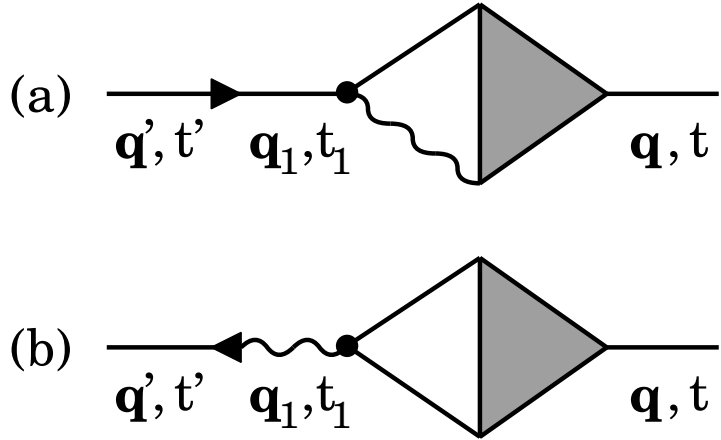}}
\centerline{\parbox{12.0cm}{\caption{\protect\small{
Block diagrams for the correlation function $C({\bf q},t,t')$. (a)
Incoming correlation propagator ${\cal C}_0({\bf q}_1,t_1;{\bf
q}',t')$. (b) Incoming response propagator ${\cal G}_0({\bf
q}',t';{\bf q}_1,t_1)$ (see \protect\Eq{GpCp} and main text). The
shaded triangle has the same meaning as in Fig.\protect\ref{Gblck}.}}
\label{Cblck}}}
\end{figure}

In contrast to Fig.\ref{Gblck} the perturbative contributions to the
correlation function $C({\bf q},t,t')$ cannot be represented by a
single block diagram. Instead, two types of block diagrams are required
as shown in Fig.\ref{Cblck}. Due to the initial condition $h({\bf
q},0)=0$ both block diagrams vanish identically for $t'=0$.
Following Ref.\cite{JSS} Fig.\ref{Cblck}
is used to obtain an exact expression for the {\em derivative} of $C$
with respect to the time argument $t'$. The diagrams for $\partial C /
\partial t'$ are of the same form as those for $C$. The main
difference between the diagrams shown in Fig.\ref{Cblck}(a) and
Fig.\ref{Cblck}(b) is that in (b) the internal time $t_1$ is
restricted to the interval $0 \leq t_1 \leq t'$ due to causality, so
that this block diagram vanishes identically for $t' = 0$. The
remaining block diagram (a) is of the same type as the block diagram
for the response function $G$ shown in Fig.\ref{Gblck}. One therefore
has a {\em termwise} correspondence between the perturbation series
for $\partial C({\bf q},t,t')/\partial t'|_{t'=0}$ and $G({\bf
q},t,t'=0)$. Gaussian theory (see \Eq{GC}) yields
\begin{equation} \label{C0tG0}
{\partial \over \partial t'}\, C_0({\bf q},t,t'=0) =
2D\, e^{-\nu {\bf q}^2 t} = 2D\, G_0({\bf q},t,t'=0)
\end{equation}
and Figs.\ref{Gblck} and \ref{Cblck}(a) then show that the two
perturbation series only differ by an overall factor $2D$. Therefore
\Eq{C0tG0} implies the relation
\begin{equation} \label{CtG}
{\partial \over \partial t'}\, C({\bf q},t,t'=0) = 2D\, G({\bf q},t,t'=0)
\end{equation}
between the correlation function $C$ and the reponse function $G$ of
the KPZ equation order by order in perturbation theory. According to
the standard Feynman rules the arguments presented above for $C$ and
$G$ also hold for arbitrary $n$-point correlation functions which
differ only in the propagator (response or correlation) assigned to
one of the external legs. Therefore \Eq{CtG} already establishes the
proof of the operator identity \Eq{hth} used in Sec.II. However, it
must be pointed out here that the above arguments only constitute a
rigorous proof of \Eq{CtG} and therefore of \Eq{hth} if the
renomalization group fixed point is accessible by perturbation
theory. For the KPZ equation this is only possible in $d=1$. In $d
\geq 2$ one encounters the well known strong coupling behavior which
forms a formidable obstacle for analytic theories of dynamic scaling
of \Eq{KPZeq} \cite{FT,FTH}. For the above derivation this means that
\Eq{CtG} may not hold in $d \geq 2$ at the renormalization group fixed
point despite its validity to all orders in perturbation
theory. Nonetheless the above perturbative analysis provides some
evidence that \Eq{CtG} and therefore \Eq{hth} holds beyond $d = 1$.

For \Eq{idMBE} the building blocks of the perturbation theory can
again be taken from Fig.\ref{GCV}, with the modification that the
response and the correlation propagator (see \Eq{GpCp}) are now given
by \Eq{GCMBE} and that the expression $\lambda_1 {\bf q}_1\cdot{\bf
q}_2 ({\bf q}_1+{\bf q}_2)^2$ must be assigned to each vertex as can
be read off from \Eq{Jhh1MBE}. It is then straightforward to see that
the arguments given above for the KPZ equation can be directly applied
to ideal MBE dynamics, where no strong coupling behavior is
encountered in any spatial dimension of physical interest. The exact
relations given by \Eqs{Gq0}{CtG} and the operator identity \Eq{hth}
therefore also hold for \Eq{idMBE}.

\section{Response and correlation functions}
 \setcounter{equation}{0}
In order to justify \Eqs{Z}{Z1} within the dimensional regularization
scheme \cite{FT} in the ${\bf q},t$ representation and to obtain
some indication how the short-time to long-time crossover takes
place the response and correlation functions of \Eqs{KPZeq}{idMBE} are
calculated here to one-loop order.

The one-loop contribution to the response function for the KPZ
equation is given by the block diagram shown in Fig.\ref{Gblck}, where
the shaded triangle is replaced by a single vertex shown in
Fig.\ref{GCV}. The analytic expression for this diagram is then given
by
\begin{eqnarray} \label{G1}
G_1({\bf q},t,t') &=& \lambda^2 \int_0^\infty dt_1 \int_0^\infty dt_2
\int {d^dq' \over (2\pi)^d} ({\bf q}'\cdot {\bf q})({\bf q}'\cdot
({\bf q}' - {\bf q})) \\
&\times& G_0({\bf q},t_1,t') G_0({\bf q}'-{\bf q},t_2,t_1)
C_0({\bf q}',t_2,t_1) G_0({\bf q},t,t_2), \nonumber
\end{eqnarray}
where $G_0$ and $C_0$ are given by \Eq{GC}. For simplicity we only
consider \Eq{G1} in the limit ${\bf q} \to {\bf 0}$ so that we can
employ the expansion
\begin{equation} \label{G0q0}
G_0({\bf q}'-{\bf q},t_2,t_1) = G_0({\bf q}',t_2,t_1) \left[1 + 2\nu
({\bf q}'\cdot {\bf q}) (t_2 - t_1) + {\cal O}({\bf q}^2) \right].
\end{equation}
The ${\bf q}'$ integration in \Eq{G1} to leading order in ${\bf q}$
is then reduced to the calculation of second moments of a Gaussian
in $d$ dimensions. The result is
\begin{eqnarray} \label{G1intq}
G_1({\bf q},t,t') &=& {\bf q}^2 {g \over 2^d \pi^{d/2}}\
G_0({\bf q},t,t') (2\nu)^{2-d/2} \\
&\times& \int_{t'}^t dt_2 \int_{t'}^{t_2} dt_1 \left[{d-2 \over 2 d}
(t_2-t_1)^{-d/2} -\left({d-2 \over 2 d}-{t_1 \over 2t_2}\right)
t_2^{-d/2} \right] , \nonumber
\end{eqnarray}
where the effective coupling constant $g$ is defined by \Eq{g}. The
remaining integrals in \Eq{G1intq} can be easily performed using
dimensional regularization \cite{FT} with $d = 2 + \varepsilon$ in the
exponents of $t_2-t_1$ and $t_2$. Note, that the prefactor $d-2$ in
\Eq{G1intq} comes from an angular integration and must not be
cancelled by factors $1/\varepsilon$ indicating UV singularities
in the time integral \cite{FT}. With the definition of $u$
according to \Eq{Rpar} ($Z_g = 1$ at this order) one obtains for
$G({\bf q},t,t')=G_0({\bf q},t,t')+G_1({\bf q},t,t')$
\begin{equation} \label{G}
G({\bf q},t,t') = G_0({\bf q},t,t')
\left\{1 - {{\bf q}^2 \over 2} u \mu^{-\varepsilon}
\left[{d-2 \over d \varepsilon} (2\nu (t-t'))^{2-d/2} +
{d-4 \over 4 d} {(2\nu (t-t'))^2 \over (2\nu t)^{d/2}} \right]\right\}.
\end{equation}
The $1/\varepsilon$ pole (the UV singularity) in \Eq{G} can be
removed, e.g., by requiring $G({\bf q},t^R,0)$ to stay finite for
$\varepsilon \to 0$, where $t^R \equiv 1/(2\mu^2 \nu^R)$ is a
reference time and $\nu^R$ is given by \Eq{Rpar}. Minimal subtraction
yields the renormalization factor $Z_\nu$ quoted in \Eq{Z}. Note that
the short-time contribution to $G$ does not produce an additional
$1/\varepsilon$ pole. By naively exponentiating the ${\bf
q}$-dependence of $G$ in the long-time limit one obtains at the
infrared stable fixed point $u = u^* \neq 0$
\begin{equation} \label{GR}
G^R({\bf q},t,t') = \Theta(t-t') \exp \left[-{{\bf q}^2 \over 2\mu^2}
\left( 2\nu^R \mu^2 (t-t') \right)^{2/z}\right]
\left[1 - u^* {{\bf q}^2 \over \mu^2} {d-4 \over 8 d}
{\left(2\nu^R \mu^2(t-t')\right)^2 \over \left(2\nu^R \mu^2 t
\right)^{d/2}} \right] .
\end{equation}
The predictive value of \Eq{GR} is very limited because $u^*$ is
{\em infinite} for $d \geq 2$. In $d = 1$ \Eq{GR} indicates that the
combination $(t-t')^2/t^{d/2}$ of the time arguments governs the
crossover to the long-time scaling behavior of $G$ for $t \to \infty$
with {\em fixed} $t-t'$. From dimensional arguments and the fact that
the short-time contribution to $G$ does not produce additional UV
singularities we can infer that according to \Eq{GR} 
${\bf q}^2 (t-t')^2/t^{d/2}$ for $d = 1$ is the scaling argument which
governs the leading finite-time correction to the
asymptotic long-time behavior of $G$. Furthermore, \Eq{GR} shows that
$G^R$ is analytic in $t'$ for $t' \ll t$ at the one-loop level, but
this behavior may be modified in higher orders. Finally, we note that
the scaling form of the asymptotic long-time contribution to $G^R({\bf
q},t,t')$ given by the exponential in \Eq{GR} has recently been
derived by combining perturbative methods with a mode coupling theory
for the KPZ equation \cite{FTH}.

The correlation function $C({\bf q},t,t')$ for \Eq{KPZeq} can be
discussed in much the same way as the response function. This time we
simplify the calculations even further by limiting ourselves to ${\bf
q} = {\bf 0}$. In this case only the diagram in Fig.\ref{Cblck}(b)
contributes and we obtain to one-loop order
\begin{equation} \label{C1}
C({\bf 0},t,t') = 2D \min(t,t') + {\lambda^2 \over 2} \int_0^t dt_2
\int_0^{t'} dt_1 \int {d^d q \over (2\pi)^d} q^4
\left[C_0({\bf q},t_1,t_2)\right]^2 ,
\end{equation}
where $C_0$ is given by \Eq{GC}. The integrations in \Eq{C1} can be
easily performed and using dimensional regularization one arrives at
\begin{eqnarray} \label{C}
C({\bf 0},t,t'\leq t) &=& 2D\left\{t' + {u \mu^{-\varepsilon} \over 4\nu
\varepsilon} \left[(2\nu(t-t'))^{2-d/2} - (2\nu(t+t'))^{2-d/2}
\right. \right. \\
&+& \left. \left. (2\nu t)^{2-d/2} \left((4-d)
(t'/ t) - d (t'/ t)^{2-d/2}\right) \right] \right\} , \nonumber
\end{eqnarray}
where \Eq{Rpar} has been used with $Z_g = 1$.
The $1/\varepsilon$ pole in \Eq{C} can be removed by demanding $C({\bf
0},t^R,t^R) = \mbox{finite}$, where $t^R \equiv 1/(4\mu^2 \nu^R)$ is
chosen as the reference time. Using minimal subtraction one finds
the renormalization factor $Z_D$ quoted in \Eq{Z}. For $t' \ll t$
\Eq{C} can be simplified to
\begin{equation} \label{Cshort}
C({\bf 0},t,t' \ll t) = 2Dt' \left[1-{u \over \varepsilon} {d \over 2}
(2\nu \mu^2 t')^{1-d/2} + {\cal O}(t'^2) \right]
\end{equation}
which explicitly shows that the short-time contribution to $C$
produces an additional $1/\varepsilon$ pole. For $t' > 0$ the
renormalized correlation function can be naively exponentiated at the
infrared stable fixed point $u = u^* \neq 0$. The result is
\begin{eqnarray} \label{CR}
C^R({\bf 0},t,t' \leq t) &=& D^R \left[
(t+t') \left(2\nu^R \mu^2 (t+t')\right)^{\theta-1} -
(t-t') \left(2\nu^R \mu^2 (t-t')\right)^{\theta-1} \right. \\
&-& \left. 2\theta t' \left(2\nu^R \mu^2 t\right)^{\theta-1} +
2t' \left(2\nu^R \mu^2 t'\right)^{\theta-1} \right] , \nonumber
\end{eqnarray}
where $\theta$ is the short-time exponent given by \Eq{theta} and
$d=1$ has been assumed. The short-time scaling behavior for $t' \ll t$
is also reproduced by \Eq{CR}. However, from \Eqs{hth}{Gq0} one
expects ${\partial \over \partial t'} C^R({\bf 0},t,t'=0) = 2D^R$
which is {\em not} reproduced by \Eq{CR}, because $\theta > 1$.
Therefore \Eq{CR} can only give a rough idea of the true scaling form
of the correlation function $C^R$ for the KPZ equation. However,
\Eq{CR} indicates, that for ${\bf q}={\bf 0}$ short-time corrections
to the correlation function persist indefinitly (see also \Eq{C}).

For ideal MBE dynamics according to
\Eq{idMBE} the one-loop contribution to the response function
is again given by the block diagram shown in Fig.\ref{Gblck}, where
the shaded triangle is replaced by a single vertex. The analytic
expression for this diagram is then given by
\begin{eqnarray} \label{G1MBE}
G_1({\bf q},t,t') &=& 4\lambda_1^2 {\bf q}^2 \int_0^\infty dt_1
\int_0^\infty dt_2 \int {d^dq' \over (2\pi)^d} ({\bf q}'\cdot {\bf
q})({\bf q}'\cdot ({\bf q}' - {\bf q})) ({\bf q}'-{\bf q})^2\\
&\times& G_0({\bf q},t_1,t') G_0({\bf q}'-{\bf q},t_2,t_1)
C_0({\bf q}',t_2,t_1) G_0({\bf q},t,t_2), \nonumber
\end{eqnarray}
where $G_0$ and $C_0$ are given by \Eq{GCMBE}. For simplicity we only
consider \Eq{G1MBE} in the limit ${\bf q} \to {\bf 0}$, i.e., we use
the expansion
\begin{equation} \label{G0q0MBE}
G_0({\bf q}'-{\bf q},t_2,t_1) = G_0({\bf q}',t_2,t_1) \left[1 + 4\nu_1
{\bf q}'^2({\bf q}'\cdot {\bf q}) (t_2 - t_1) + {\cal O}({\bf q}^2) \right].
\end{equation}
The ${\bf q}'$ integration in \Eq{G1MBE} to leading order in ${\bf q}$
yields
\begin{eqnarray} \label{G1MBEintq}
G_1({\bf q},t,t') &=& {q^4 \over 4} {g_1 \over 2^d \pi^{d/2}}\
{\Gamma(d/4) \over \Gamma(d/2)}\ G_0({\bf q},t,t') (2\nu_1)^{2-d/4} \\
&\times& \int_{t'}^t dt_2 \int_{t'}^{t_2} dt_1 \left[{d-6 \over d}
(t_2-t_1)^{-d/4} -\left({d-6 \over d}-{t_1 \over t_2}\right)
t_2^{-d/4} \right] , \nonumber
\end{eqnarray}
where the effective coupling constant $g_1$ is defined by \Eq{g1}. As
in \Eq{G1intq} the remaining integrals in \Eq{G1MBEintq} can be
performed using dimensional regularization with $d = 4 + \varepsilon$
in the exponents of $t_2-t_1$ and $t_2$. As usual the $1/\varepsilon$
poles indicate UV singularities in the time integral. With
the definition of $u$ according to \Eq{Rpar1} and $Z_{g_1} = 1$ one
obtains for $G({\bf q},t,t')=G_0({\bf q},t,t')+G_1({\bf q},t,t')$
in the limit $t \to \infty$ with $t-t' = \mbox{const}$
\begin{equation} \label{GMBE}
G({\bf q},t,t') = G_0({\bf q},t,t')
\left\{1 - {q^4 \over 2} u \mu^{-\varepsilon}
\left[{d-6 \over d \varepsilon} (2\nu_1 (t-t'))^{2-d/4} + {3 \over 4}\
{d-8 \over 4 d} {(2\nu_1(t-t'))^2 \over (2\nu_1t)^{d/4}} \right]\right\}
\end{equation}
up to terms ${\cal O}((t-t')^3/t^{d/4+1})$.
The $1/\varepsilon$ pole (the UV singularity) in \Eq{GMBE} can be
removed by the minimal subtraction scheme described above, where
$t^R \equiv 1/(2\mu^4 \nu_1^R)$ defines the reference time and
$\nu_1^R$ is given by \Eq{Rpar1}. One obtains the renormalization
factor $Z_{\nu_1}$ quoted in \Eq{Z1}. By naively exponentiating the
${\bf q}$-dependence of $G$ in the long-time limit one obtains at the
infrared stable fixed point $u = u^* \neq 0$
\begin{equation} \label{GRMBE}
G^R({\bf q},t,t') = \Theta(t-t') \exp \left[-{q^4 \over 2\mu^4}
\left( 2\nu_1^R \mu^4 (t-t') \right)^{4/z}\right]
\left[1 - {3 \over 4} u^* {q^4 \over \mu^4} {d-8 \over 8 d}
{\left(2\nu_1^R \mu^4(t-t')\right)^2 \over \left(2\nu_1^R \mu^4 t
\right)^{d/4}} \right] .
\end{equation}
In contrast to \Eq{KPZeq} the infrared stable fixed point for
\Eq{idMBE} is {\em finite} in any dimension of physical interest.
Especially one has $u^* = {\cal O}(\varepsilon)$ so that an
$\varepsilon$-expansion around the upper critical dimension $d_c=4$
can be performed. The qualitative behavior of $G^R$ according to
\Eq{GRMBE} is very similar to the behavior of $G^R$ for the KPZ
equation in $d=1$ (see \Eq{GR}). Here the leading finite-time
correction to the asymptotic long-time behavior is governed by the
combination $(t-t')^2/t^{d/4}$ of time arguments.

The one-loop contribution to the correlation function $C({\bf q}={\bf
0},t,t')$ for \Eq{idMBE} vanishes identically due to an additional
factor ${\bf q}^2$ in the vertex (see \Eq{G1MBE}) so that
\begin{equation} \label{CRMBE}
C^R({\bf 0},t,t') = 2D \min(t,t') + {\cal O}(u^2) .
\end{equation}
\Eq{CRMBE} directly demonstrates that $Z_D = 1$ as quoted in \Eq{Z1} to
one-loop order. As in the case of KPZ dynamics \Eq{CRMBE} demonstrates
that finite-time corrections to the correlation function for
\Eq{idMBE} persist indefinitely for ${\bf q} = {\bf 0}$.

\end{document}